\documentstyle[amssymb,aps,twocolumn]{revtex}

\begin{document}
\title{Ambipolar gate effect and low temperature magnetoresistance of ultrathin La$%
_{0.8}$Ca$_{0.2}$MnO$_{3}$Films}
\author{M. Eblen-Zayas, A. Bhattacharya, N. E. Staley, A. L. Kobrinskii, and A. M.
Goldman}
\address{School of Physics and Astronomy, University of Minnesota, 116 Church St. SE,%
\\
Minneapolis, MN 55455, USA}
\date{
\today%
}
\maketitle

\begin{abstract}
Ultrathin La$_{0.8}$Ca$_{0.2}$MnO$_{3}$ films have been measured in a
field-effect geometry. The electric field due to the gate produces a large
ambipolar decrease in resistance at low temperatures. This is attributed to
the development of a pseudogap in the density of states and the coupling of
localized charge to strain. The gate effect and magnetoresitance are
interpreted in a consistent framework. The implications for the low
temperature behavior of a manganite film in the two dimensional limit are
discussed.
\end{abstract}

\pacs{PACS numbers: ()}

Ultrathin films of manganites present a unique opportunity for studying the
low temperature properties of a half-metallic fully spin polarized electron
gas in the two-dimensional limit, with tunable disorder. In the past decade,
there has been a resurgence of interest in the possible ground states of
highly correlated two-dimensional electron gases (2DEGs) \cite{abrahams},
especially regarding the existence of a metallic state at low temperatures.
Both high mobility 2DEGs and highly disordered amorphous ultrathin metal
films exhibit large positive magneto-resistance for in-plane magnetic fields
at low temperatures \cite{mertes}. This is consistent with theoretical
arguments that attribute the existence of a metallic ground state to strong
antiferromagnetic or singlet correlations between electrons \cite{phillips}
in the appropriate range of densities. When these correlations are quenched
by the Zeeman energy in a magnetic field, the metallic state is lost, and
the resulting insulator accounts for the large positive magnetoresistance.
In the manganites, due to strong Hund's coupling between the core Mn spins
and the conduction electrons, a fully spin-polarized electron gas is
obtained at low temperatures in the ferromagnetic state. This does not allow
for singlet correlations, and should result in an insulating ground state in
the two-dimensional limit, even in the presence of very weak disorder.
Furthermore, the strong cross-couplings between strain, charge and spin that
exist in these systems make for unique features in their response to
external perturbations. Localized electrons at the Mn$^{3+}$ sites cause
lattice distortions via the Jahn-Teller effect, and as a result, strain
fields develop within the lattice. When electrons are delocalized, either by
better alignment of spins by an external magnetic field (double-exchange
mechanism) or via changing the local density of charge (electrostatic
gating), the strain is relieved and the resulting disorder potential is
reduced.

In this letter we investigate the low temperature properties of ultrathin
films of La$_{0.8}$Ca$_{0.2}$MnO$_{3}$ (LCMO) to gate and magnetic fields.
This composition of LCMO is close to the phase boundary between a
ferromagnetic metal (FM) at higher Ca doping and a ferromagnetic charge
ordered insulator (F-COI) at lower doping. Bulk single crystals of similar
composition are believed to exist in a mixed phase with coexisting regions
of insulating and metallic properties \cite{algarabel}, with the behavior at
low temperatures arising from percolation of metallic regions in the
material. The samples are typically 21 u.c ($\backsim $82\AA ) thick films
of La$_{0.8}$Ca$_{0.2}$MnO$_{3}$ grown using ozone-assisted molecular beam
epitaxy on surface treated SrTiO$_{3}$ substrates that have been thinned to
35-50$\mu $m locally\cite{bhatt1} permitting configuration in a field-effect
geometry. A Pt electrode (1000\AA\ thick) deposited on the back of this
thinned region serves as the gate. The as-grown manganite film is patterned
into a wire of 100$\mu $m width, with tabs for carrying out four terminal
measurements. All transport measurements reported here were performed using
standard DC techniques. The gate-drain current was always monitored, and
remained below 0.6nA for gate electric fields as high as 85kV/cm, while
typically the source-drain measurement current was 100nA.

We observe a magnetic transition at about 150K with an accompanying
resistive transition at about 138K (T$_{c}$), from an activated insulating
state to a nominally metallic state, along with CMR [Fig.1(a)]. However, at
the lowest temperatures (%
\mbox{$<$}%
36K) there is a re-entrant insulating phase \cite{ziese}. Near the resulting
minimum, the film resistance has a large ambipolar susceptibility to a gate
electric field along with a large magnetoresistance, in contrast to earlier
work on thicker films \cite{tabata} where the largest susceptibilities were
observed only near the insulator-metal transition at high temperature. We
propose that the ambipolar gate response is due to the opening up of a
pseudogap in the density of states (DOS) at low temperatures. The large
magnitude of the response to gate and magnetic fields is due to the motion
of domain boundaries separating insulating and metallic phases. Glassy
dynamics have also been observed in the response of these films to external
perturbations \cite{bhatt2}, attributed to the relaxation of strain via
cross-couplings to spin and charge.

The transition from the low temperature metal to the re-entrant insulator at
the lowest temperatures is continuous. This is borne out by the absence of
any hysteresis upon cooling and then warming the sample, for all
temperatures below 30K. There is hysteresis above this temperature,
associated with the insulator-metal transition at T$_{c}$, as seen in other
work \cite{chen}. We interpret the hysteretic behavior as a signature of the
nucleation (melting) of metallic patches upon cooling (warming) near T$_{c}$%
. Upon cooling to low temperatures, these patches form an infinite cluster
and metallic behavior ensues. Within that framework, the lack of hysteresis
in the transition into the re-entrant insulator indicates that the
insulator/metallic fraction does not change upon entering this phase.
However, the degree of disorder (strain) within the insulating patches, and
the resultant martensitic accommodation strain may become higher, as has
been observed in optical images at low temperatures \cite{podzorov}.

It is known from photoemission\cite{park} and optical conductivity\cite
{okimoto} experiments that the density of states (DOS) at the Fermi level in
the manganites is severely depleted, presumably due to the opening of a
pseudogap \cite{saitoh} or a minimum in the DOS at the chemical potential.
This has also been observed in scanning tunneling spectroscopy at low
temperatures in the mixed phase \cite{fath}. A pseudogap is predicted in the
mixed phase in two dimensions in the presence of either structural \cite
{moreo} or spin disorder \cite{kumar}. The minimum is most pronounced at low
temperatures, and is eventually washed out as the temperature is raised. A
pseudogap has been observed in photoemission experiments on layered
manganites \cite{saitoh}, with an accompanying upturn in the resistance at
the lowest temperatures.

We studied the response of the film to an applied gate field in the
temperature range from 2K to 138K [Fig.2(a)]. The most striking feature of
the data is that the response is ambipolar for nearly the whole temperature
range. At T$_{c}$, the resistance decreases for negative gate voltages
(which induce hole type carriers), but there is no change in resistance for
positive gate voltages. Upon lowering the temperature, the response to a
positive gate voltage grows, although there is a larger effect for negative
gate voltages than for positive. The ambipolar nature of the response may
suggest electrostriction in the STO being the cause, but several results
refute this. LCMO thin films exhibit an increase in resistance, upon an
increase in the strain (positive sign for biaxial expansion) \cite{rao}. The
maximum strain due to electrostriction \cite{bhatt1} is about 6x10$^{-4}$,
almost an order of magnitude less than the strain due to the lattice
mismatch between STO and LCMO, which is 7.7x10$^{-3}$. The positive sign of
the strain indicates that electrostriction would increase the mismatch
between LCMO and STO, increasing the resistance further, and not decreasing
it as we observe. Additionally, STO exhibits no appreciable electrostriction
effects below electric fields of 2kV/cm, and the effect saturates at
electric fields above 15kV/cm \cite{bhatt1} at low temperatures. We find
electric fields as low as 0.29 kV/cm reduce the resistance, and there is no
evidence of saturation of the gate effect for electric fields as high as
85kV/cm. Furthermore, using an identical dielectric, non-ambipolar
modulation of the transition temperature has been obtained for ultrathin NdBa%
$_{2}$Cu$_{3}$O$_{7-\delta }$ films in a field effect geometry \cite{matthey}%
, with enhancement (reduction) of T$_{c}$ with a negative (positive) gate
voltage. Thus is likely that the ambipolar effect reflects the electronic
properties of the ultrathin LCMO.

An ambipolar gate effect can be signature of a minimum in the density of
states. In amorphous covalent semiconductors, this occurs due to tails in
the DOS in the mobility gap \cite{barbe}. In a two-dimensional disordered
system, the evolution from a metal to an insulator (e.g. as a function of
decreasing thickness) may be accompanied by the opening of a correlation gap
in the DOS at the Fermi level \cite{butko}. Normally, one would not expect
an ambipolar gate effect if the minimum in the DOS tracks the chemical
potential. However, if the relaxation time for charge to equilibrate is
longer than the measurement time, turning on a gate voltage abruptly can
lead to a non-ergodic distribution of charge, where the chemical potential
rides up the DOS about the minimum for both signs of applied gate voltage,
giving rise to an ambipolar enhancement of conductance. This is usually
accompanied by glassy behavior in the charge relaxation \cite{yu}.

Our observations may be explained by the physics of the mixed phase at low
temperatures. The strongly insulating regions would be largely transparent
to the gate electric field, and remain unaffected. The strongly conducting
regions would have a DOS consistent with the dopant concentration, and would
also be weakly affected, since the electric field would be screened out in
the first few unit cells. The boundaries of these regions would have
intermediate DOS and would couple maximally. Because of strong coupling
between localized charge and local strain via the Jahn-Teller effect, any
change in the charge configuration in these boundary regions causes a
corresponding change in the strain. Competing strain fields in the material
gives rise to a metastable energy landscape with hierarchical energy
barriers for relieving the strain \cite{ahn}. This causes a hierarchical and
glass like response of the system to any external force, due to either gate
or magnetic field, that seeks to change the phase boundaries between the
insulating and conducting regions. The hierarchy of energy barriers being
crossed in either case is the same since they are governed by the same
strain field \cite{bhatt2}. If the DOS has a minimum, and the electronic
response is glass like, the conductance of these boundary regions will be
enhanced for both signs of gate voltage. The resulting delocalized charge
causes the strain fields to be relieved and the phase fraction of the
metallic region increases at the expense of the insulating region. Domain
boundaries move irreversibly through a metastable pinning landscape, and the
system latches the change caused in the conductance by the gate.

We first comment on the qualitative nature of the gate effect and its
temperature dependence. We use the ratio of the magnitude of the gate effect
for positive and negative voltages as a measure of ambipolarity, independent
of the magnitude of the effect. Upon lowering the temperature below T$_{c}$,
this measure is seen to increase and then level off [Fig.2(b)] at the lowest
temperatures, nominally consistent with the predicted evolution of the
pseudogap \cite{moreo,kumar}. The pseudogap is likely suppressed by the
application of a magnetic field, since this reduces the disorder and makes
the sample becomes more homogeneously metallic. This is borne out by the
suppression of the upturn by application of a magnetic field [Fig.2(c)].

Next, we turn to the magnitude of the effect. In understanding the response,
we have to consider the nonlinear and temperature dependent dielectric
constant of STO \cite{bhatt1}, and the susceptibility of the film. The
maximum response to electric field occurs in the region near the onset of
the reentrant insulating state. Although the response is greatest at 30K,
the response at 50K is comparable when considering that dielectric constant
of STO \cite{bhatt1} at 50K is less than at 30K and thus the equivalent
charge transfer at that temperature will be less. However, at 2K, when the
induced charge is maximal ($\backsim $5.8x10$^{13}$cm$^{-2}$ for 40kV/cm)
the response is not as large as at 50 K where the charge transfer is
significantly less ($\backsim $4.4x10$^{13}$cm$^{-2}$ for 40kV/cm). Other
manganite films with re-entrant insulating behavior have also shown a
reduction in the gate effect at temperatures below 50K \cite{pallecchi}.
Also, looking at the magnetoresistance data, the resistance of the manganite
film is more susceptible to the initial application of a magnetic field at
30K than at 2K [Fig.3(a)] We consider this in the framework of pinned domain
walls separating insulating and metallic regimes, where the hierarchy of
pinning is determined by strain fields, which can be coupled to both by
magnetic and electric fields \cite{bhatt2}, one would expect the magnitude
of the response to gate electric field and magnetic field to exhibit a
similar temperature dependent response. At 30K, the gate effect is
suppressed magnetic field [Fig.3(b)]. In zero magnetic field, the gate
effect at 2K is less than at 30K [Fig.2(a)]. However, when 9T is applied to
the film and then a gate is turned on, the gate effect is larger at 2K than
at 30K [Fig 3(c)]. All of this can be understood in the framework of a
`general susceptibility', describing the ease with which a domain boundary
moves in a hierarchical pinning landscape. Since magnetic and gate electric
fields are `equivalent' in the response of the film in the mixed phase, we
use the magnetoresistance as a measure of this susceptibility. The
magnetoresistance as a function of field [Fig.3(a)] clearly shows that the
slope is steeper at 30K than at 2K for low fields. However at higher fields,
this situation is reversed. Thus, turning on a small gate field in the
presence of a magnetic field would have the effect of taking a differential
measurement of the MR, and this explains the difference between the 2K and
30K data in the presence and absence of a magnetic field.

A number of reports have been made of a field-direction independent response
to applied gate electric fields in manganite thin films \cite{tabata,ogale}.
One model for the ambipolar response proposes that the applied electric
field affects the lattice distortions in the manganite, causing
corresponding changes in resistivity \cite{ogale}. In this picture, with the
applied electric field acting as a perturbation to the Jahn-Teller
distortion within the manganite, it is unclear why the ambipolar response
would have the temperature dependence that we see. Indeed, the original
proposal was used to explain ambipolar behavior in the vicinity of and above
the temperature of the peak resistance.

Another alternative mechanism that does not involve the density of states
directly is due to just the electrostatic forces felt by the accumulated
charges in the metallic regions. As mentioned earlier, the insulating
regions are transparent to the gate electric field, and some of these field
lines would terminate on metallic regions at the boundaries with the
insulating regions. These charges would then experience a force that would
literally pull the boundaries further into the insulating regions,
increasing the fraction of metallic phase. The same arguments regarding the
hierarchy of barriers surpassed would hold. However, in a purely hole doped
system, there would also be accumulation and depletion at the boundaries due
to negative and positive gate voltages, and it has been argued that this
gives rise to a non-ambipolar effect \cite{tabata}. The temperature
dependence of the symmetry of such an effect is also not obvious.

In conclusion, the nature and magnitude of the gate effect, and the relation
to magnetoresistance are consistently explained within a mixed phase
scenario, invoking a pseudogap and a hierarchical metastable energy
landscape for the motion of the phase boundaries. The absence of an
ambipolar effect at higher temperatures is likely due to a greater
homogeneity of phases in this regime. The low dimensionality of our films
may also be crucial in observing the effects. The possibility of a field
(disorder) induced insulator-metal transition in the re-entrant insulator
with the existence of a quantum critical point will be the subject of future
work.

This work was supported by the National Science Foundation under grant
NSF/DMR-0138209 and by the University of Minnesota MRSEC (NSF/DMR- 0212032)$%
. $ M. Eblen-Zayas was supported by an NSF Graduate Research Fellowship.


\begin{figure}[tbp]
\caption[Evolution of$R(T)$ of the 9.19\AA\ film as a function of in-plane
magnetic field. Field values from top to bottom are: 12.5, 12, 11.6, 11.5,
11, 10, 9, 8, 7, 6, 5, 4, 3, 2, and 0 T. Inset: temperature at which dR/dT
becomes zero is plotted vs. applied field.]{Ambipolar gate effect and the
pseudogap. (a) Temperature dependence of the gate effect. Note, this does
not account for the temperature and field dependence of the dielectric
properties. (b) Temperature dependence of the `ambipolarity' or symmetry of
the gate response. (c) Dependence of the re-entrant phase on applied
magnetic field (different sample).}
\label{fig2}
\end{figure}
\begin{figure}[tbp]
\caption{Susceptibility to external fields. (a) Magnetoresistance for
in-plane fields. The data at 2K and 30K have greater density of points. (b)
Gate effect at different magnetic fields at 30K. (c) Gate effect at 2K and
30K, with 9T magnetic field on.}
\label{fig3}
\end{figure}

\end{document}